\documentclass[aps,prb,twocolumn,superscriptaddress,showpacs,citeautoscript]{revtex4-1}

\usepackage{color}
\usepackage{graphicx}
\usepackage{bm}

\begin{document}
\par
\title{Equilibrium intermediate-state patterns  in  a type-I superconducting slab in an arbitrarily oriented applied magnetic field
}
\author{John R. Clem}
\affiliation{Ames Laboratory - DOE and Department of Physics and
Astronomy, Iowa State University, Ames Iowa 50011, USA }

\author{Ruslan Prozorov}
\affiliation{Ames Laboratory - DOE and Department of Physics and
Astronomy, Iowa State University, Ames Iowa 50011, USA }

\author{R. J. Wijngaarden}
\affiliation{ Department of Physics and Astronomy, VU University Amsterdam, 1081 HV Amsterdam, The Netherlands}

\date{\today}

\begin{abstract}

The equilibrium topology of superconducting and normal domains in flat type-I superconductors is investigated. Important improvements with respect to previous work are: (1) the energy of the external magnetic field, as deformed by the presence of superconducting domains, is calculated in the same way for three different topologies, and (2) calculations are made for arbitrary  orientation of the applied field. A phase diagram is presented for the minimum-energy topology as a function of applied field magnitude and angle. For small (large) applied fields normal (superconducting) tubes are found, while for intermediate fields parallel domains have a lower energy.  The range of field magnitudes 
for which the superconducting-tubes structure is favored shrinks when the field is more in-plane oriented.

\end{abstract}


\maketitle

\section{Introduction\label{IntroRP}}

Studies of the intermediate-state structure in type-I superconductors have a long history beginning with the
pioneering work of Landau \cite{Landau1938,Landau1943} and continuing to the present; see, for example,  Refs.~[\onlinecite{Goldstein1996,Bokil1997,Dorsey1998,Blossey2001,Hernandez2005,Du2005,Cebers2005,Menghini2005,Prozorov2005,Jeudy2004,Choksi2004,Jeudy2006,Valko2006,Prozorov2007,Prozorov2007,Menghini2007,Prozorov2008a,Berdiyorov2009,Prozorov2009,Chudnovsky2011}]. For further background  we refer the reader to several books and reviews\cite{Shoenberg1952,Livingston1969,Tinkham1996,Poole2007,Huebener2001}. Our main interest here is a superconducting slab or oblate ellipsoid with thickness much greater than the coherence length or the London penetration depth in the intermediate state, which we assume consists of normal domains of constant magnetic flux density and superconducting domains of zero  flux density.  The macroscopic Helmholtz free energy density relative to that in the Meissner state (accounting for both  the condensation-energy and field-energy costs)  is ${\cal F}(\bm B) = B_cB/\mu_0$, where $H_c = B_c/\mu_0$ is the bulk thermodynamic critical field and $\bm B$ is the average magnetic flux density in the sample.    The corresponding $\bm H$ field is\cite{Fetter69}
$\bm H = \nabla_{\bm B} {\cal F}(\bm B)$, such that $\bm H =  H_c \hat {\bm B}$  and $\hat {\bm B} = \bm B/B$.  Because these energy contributions alone are insufficient to determine the spatial distribution of the normal and superconducting domains, in this paper we examine three idealized models of the intermediate-state magnetic structure in thick superconducting slabs, accounting for the differences in wall-energy and field-energy contributions 
(the other contributions to the free energy do not depend on the topology), and we determine which model has the lowest free energy as a function of the magnitude $H_0$ and tilt angle $\theta_0$ of the applied field. 
For the definition of $\theta_0$, see Fig. \ref{BFig}.

We find that in a perpendicular ($\theta_0 = 0$) magnetic field $H_0$ the energetically favored structures are (1)  a triangular array of normal flux tubes for relatively small $H_0$, (2)  parallel normal and superconducting domains for intermediate $H_0$, and (3) a triangular array of superconducting tubes for large $H_0$.  As the tilt angle increases, however, (1) the  triangular array of normal flux tubes is energetically favored for a somewhat wider range of $H_0$, (2)  parallel normal and superconducting domains \cite{Sharvin57,Dzyaloshinskii55} are favored for a much wider range of $H_0$, and (3) the triangular array of superconducting tubes is favored for a much smaller range of $H_0$ near $H_c$.  

Experimentally, in addition to various macroscopic and indirect techniques, magnetic flux structures in type-I superconductors were visualized on the sample surface by using a Bi wire as a magnetoresistive probe \cite{Meshkovsky1947,Meshkovsky1958}, by decoration with small diamagnetic \cite{Schawlow1956} or ferromagnetic \cite{Balashova1956,Sarma1967,Lischke1974} particles, by 
the
electron mirror technique \cite{Bostanjoglo1967}, by using 
the
magneto-optical Faraday effect \cite{Alers1957,Huebener2001,Farrell1975,Castro1999,Menghini2005,Jeudy2006,Prozorov2007}, by using miniature scanning Hall probes \cite{Bending2012,Ge2013} and, in the bulk, 
by
using polarized neutron reflectometry \cite{Nutley1994,NeutronsPb_PRB2012} and muon spin rotation \cite{Egorov2001}.

Unlike type-II superconductors where the magnetic field can  appear only in the form of single-flux-quantum Abrikosov vortices \cite{Abrikosov1957}, the mix of normal and superconducting domains in the intermediate state of type-I superconductors exhibits
diverse geometric patterns, and their shape and distribution depend sensitively on many factors, including chemical, mechanical and geometrical parameters of the studied samples \cite{Livingston1969,Huebener2001,Brandt1971,Cebers2005}, history of how magnetic fields and temperature were varied, direction of the
magnetic field with respect to the sample, and dynamical perturbations such as electric currents or ac fields. The observed patterns are often quite similar to those seen (or theoretically suggested) in a variety of other strongly correlated systems,  from various foams and froths \cite{Weaire1999,Prozorov2011}, to the results of mathematical studies of nonlinear dynamics and chaos \cite{Strogatz1994,Walgraef1997}, to chemical reactions, \cite{Walgraef1997} magnetic films, \cite{Garel82,Portman2003} and the astrophysics of neutron stars \cite{Buckley2004,Metlitski2005}, all of which can be regarded as manifestations of modulated phases with competing interactions.\cite{Seul95}
To reflect these similarities tubular patterns in type-I superconductors have been called the ``suprafroth'' \cite{Prozorov2008a,Prozorov2011}. 
The type-I superconductor represents an ideal system where uncontrolled coarsening in time is replaced by a controlled coarsening in a magnetic field and there is no ``drainage'' in the S/N walls.

\begin{figure}[tb]
\begin{center}
\includegraphics[width=0.9\linewidth]{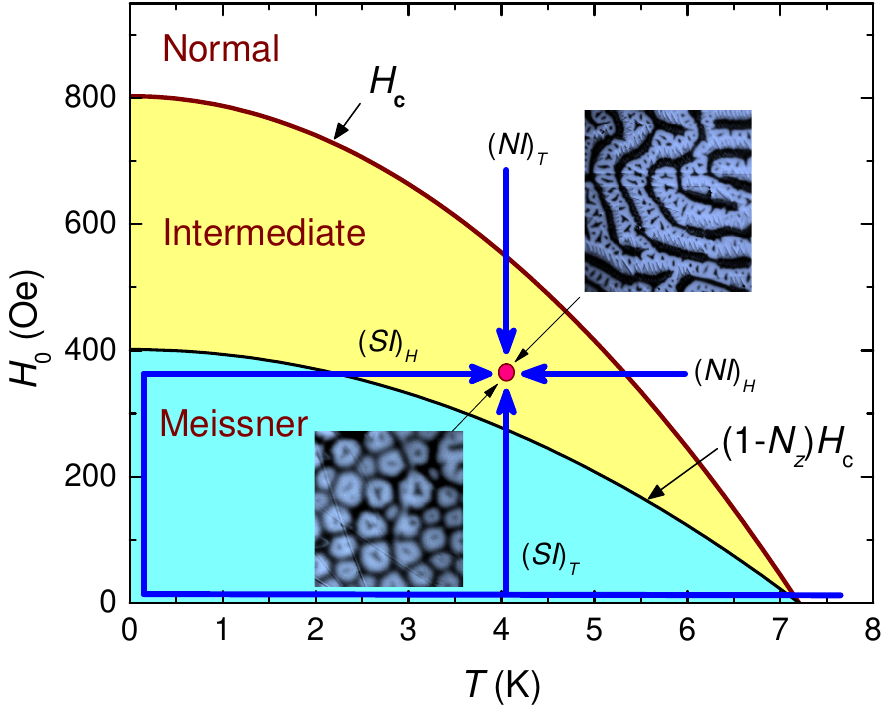}
\end{center}
\caption{(Color online) Schematic diagram of path-dependent patterns of the intermediate state in pure lead  
for perpendicularly applied magnetic fields.
\cite{Prozorov2007,Prozorov2005} 
Magneto-optical images of tubular and laminar structure are obtained at the same point on the $H-T$ phase diagram. The lower image was obtained after zero-field cooling (ZFC) and  applying the magnetic field, $(SI)_T$, and the upper image after field cooling, $(NI)_T$.}
\label{diagram}
\end{figure}

Figure \ref{diagram} outlines the schematics of four distinct ways to arrive at the same point $(H_0,T)$ with $H_0$ a perpendicular applied field and $T$ the temperature.
 The intermediate state develops above the line $(1-N_z)H_c$, where $N_z$ is the demagnetization factor,
and persists up to the critical field, $H_c$. Magneto-optical photographs show intermediate-state patterns obtained along the $(SI)_T$ path (lower left in Fig.~\ref{diagram}) and along the $(NI)_T$ path (upper right) representing flux-tubular and laminar structures, respectively.

Theoretically, the problem of the intermediate state is difficult because of multiple contributions to the free energy, the interactions inside and outside the specimen, and various additional effects that altogether determine the geometric structure of the final pattern.
Already early images of the intermediate state revealed a variety of phenomena not predicted by the simple Landau theory \cite{Meshkovsky1947,Meshkovsky1958,Shoenberg1952,Farrell1972,Chien1974}. Subsequent work found even more diverse patterns \cite{Livingston1969,Huebener2001,Menghini2007,Prozorov2008a}. In response, the initial models were refined to include domain branching \cite{Landau1938,AndrewIII,Dzyaloshinskii1956,Livingston1969,Kirchner1971a}, corrugation, \cite{Livingston1969,Kirchner1971a} and crystalline anisotropy \cite{Schawlow1958}. It seems that the once popular branching model cannot adequately account for the tubular
structure \cite{Livingston1969,Fortini1972,Kiendl1974,Huebener2001}. Several alternative  approaches have been suggested, such as flux-tube models \cite{Goren1971,Buck1981,Bokil1997}, corrections to the surface tension \cite{Mishonov1990}, higher-order expansions of the Ginzburg-Landau functional \cite{Paramos2003}, the current-loop model, \cite{Goldstein1996} and even more general thermodynamic treatments of the energy minimization problem \cite{Du2005,Choksi2004,Kozhevnikov2012}.

It is important to realize that the formation of a particular pattern strongly depends on the magnetic and thermal history of the sample. Quantized flux tubes are usually produced upon changing the  magnetic field either in magnitude or direction (or by applying an ac field) \cite{Solomon1969,Huebener1974,Huebener2001,Kirchner1972} or in the presence of an electric current \cite{Watson1974,Menghini2005,Prozorov2009}. On the other hand, when a magnetic field is held constant and there is no electric current, the pattern, at least at  intermediate fields, is lamellar-labyrinth-like. Usually, the accompanying magnetic hysteresis has been attributed to impurities, grain boundaries, dislocations, and edge barriers \cite{Livingston1969,Huebener2001}. However, recent studies have shown that this residual hysteresis remains even in the most carefully prepared samples without any bulk pinning. This so-called ``topological hysteresis'' arises from the way the intermediate state is formed after ZFC \cite{Prozorov2005,Prozorov2007,Thakur2010}. When a 
perpendicular
magnetic field is applied to a superconducting sample, it starts to protrude into the interior in fields above $(1-N_z)H_{c}$ in the form of a finger-like pattern \cite{Huebener1974,Huebener2001,Fortini1976,Clem73}. Meissner currents pinch off the protrusions in the form of flux tubes and the Lorentz force drives the tubes into the sample.  This effect is related to the so-called geometric barrier in type-II superconductors \cite{Benkraouda96}, but it has also been studied in type-I materials \cite{Castro1999}. The tubes repel each other and do not merge all way up to the normal state. This repulsion has been experimentally studied in Pb
samples \cite{Fietz1984}. 

In contrast to this behavior, upon  field-cooling from the normal state in a perpendicular magnetic field, a laminar pattern is formed initially except at low fields. 
Field-shaking experiments, however, show that the laminae that are formed upon field-cooling in high fields transform  into arrays of superconducting tubes,\cite{Menghini2007} showing that the latter structure is the equilibrium state.

For fields with arbitrary orientation with respect to the sample normal, so-called inclined fields, the phenomenology is even richer. For example, with nearly in-plane applied field, straight laminae with orientation parallel to the field are usually observed. However, when the field is subsequently tilted in the direction of the normal, corrugations start to appear. Clearly, these corrugations are not due to sample inhomogeneities, but are an intrinsic effect related to minimization of the external field energy \cite{Faber58}. Quite remarkable is the behavior for a nearly in-plane field of fixed magnitude and fixed angle with respect to the sample normal when its in-plane component is slowly rotating. Energetically, it is favorable for the long laminae to orient parallel to the (rotating) field. On the other hand, this implies very large motions of the whole laminae system. In fact, as predicted by Dorsey and Goldstein, \cite{Dorsey1998} it is observed \cite{Menghini2005} that for lead samples without pinning, a chevron phase is formed, with the laminae roughly at equal positive and negative angles with respect to the in-plane component of the applied field. On the other hand, if the lead sample does have some pinning, it is observed \cite{Menghini2005} that the laminae break into short strips that co-rotate with the applied in-plane component with a small backlag angle.

To bring out the essential physics of the intermediate state, in this paper we consider theoretically the magnetic structure that appears in a flat isotropic  type-I superconducting sample whose thickness is much smaller than its lateral dimensions. Because experiments are always done on samples of finite size, we begin Sec.\ \ref{Maxwell} with a brief discussion of demagnetization effects. 
 For simplicity we assume that magnetic flux enters the superconductor in the form of straight normal domains containing magnetic flux of density $B_c = \mu_0 H_c$, where $H_c$ is the bulk thermodynamic critical field. 

We assume that the penetration depth $\lambda$, coherence length $\xi$, and wall-energy parameter $\delta$  are much smaller than all the linear dimensions of any domain.
We do not account for the possibility that the magnitude of the flux density in the domains can differ from $B_c$ nor  that the normal-superconducting interfaces can bend near the sample surface.
Also, we do not account for corrugations of the normal-superconducting interface.\cite{Faber58}
On the other hand we assume that the slab is thin enough \cite{deGennes} so that domain branching \cite{Landau1938} does not occur.
For simplicity, we only
deal with isotropic materials for which the wall-energy parameter $\delta$ is the same for all orientations of the normal-superconducting interfaces and does not depend upon crystal-lattice effects. (For example, we do not consider the problem of anisotropic type-I superconductors, in which the normal domains tend to align themselves along certain crystal-lattice symmetry directions.)  With these assumptions, the boundary conditions on Maxwell's equations give us all the equations we need to calculate the magnitudes and directions of the average magnetic flux density $\bm B$ and the magnetic field $\bm H$ in flat isotropic  type-I superconducting sample as a function of the magnitude and direction of an applied magnetic field $\bm H_0$.

In Sec.\ \ref{IS}, we consider three separate models of the intermediate state, first in a perpendicular applied field $H_0$ and then in a magnetic field $\bm H_0$ of arbitrary angle $\theta_0$ relative to the sample normal. Using the same approach to calculate the free energies, we identify which of the three models is energetically favored in low, medium, and high fields, and we discuss how the ranges of energy favorability are affected by the field angle $\theta_0$.  Finally, we discuss our conclusions in Sec.\ \ref{Concl}.

\section{$\bm  B$, $\bm H$, $\bm M$, and Demagnetization in oblate ellipsoids and flat slabs  \label{Maxwell}}

Although experiments are always done with type-I superconductors of finite dimensions, theoretically it is often a good approximation to consider flat samples of finite thickness $d$ but infinite lateral dimensions.  To relate the two geometries we briefly discuss demagnetization effects.\cite{Cape67,Goldfarb92}  As a model sample of finite dimensions, we consider an oblate ellipsoid of revolution about the $z$ axis for which the demagnetizing factor is \cite{Goldfarb92}
\begin{equation}
N_z=(1-\gamma^2)^{-1}[1-\gamma(1-\gamma^2)^{-1/2}\cos^{-1}\gamma],
\end{equation}
where $\gamma < 1$ is the ratio of the polar axis to the equatorial axis, and $N_z \to 1- \pi \gamma/2$  in the limit as $\gamma \to 0$.  

If the superconductor is initially in the Meissner state and a magnetic field $H_0$ is applied along the $z$ axis, the field at the equator is $H_0/(1-N_z)$. Magnetic flux first penetrates there when this field is $H_c$ or $H_0 = H_c(1-N_z).$  This first-penetration field is very small (much less than $H_c$) for very thin samples, and for $H_0$ exceeding the first-penetration field up to $H_c$, the sample is in the intermediate state.

\begin{figure}
\includegraphics[width=6cm]{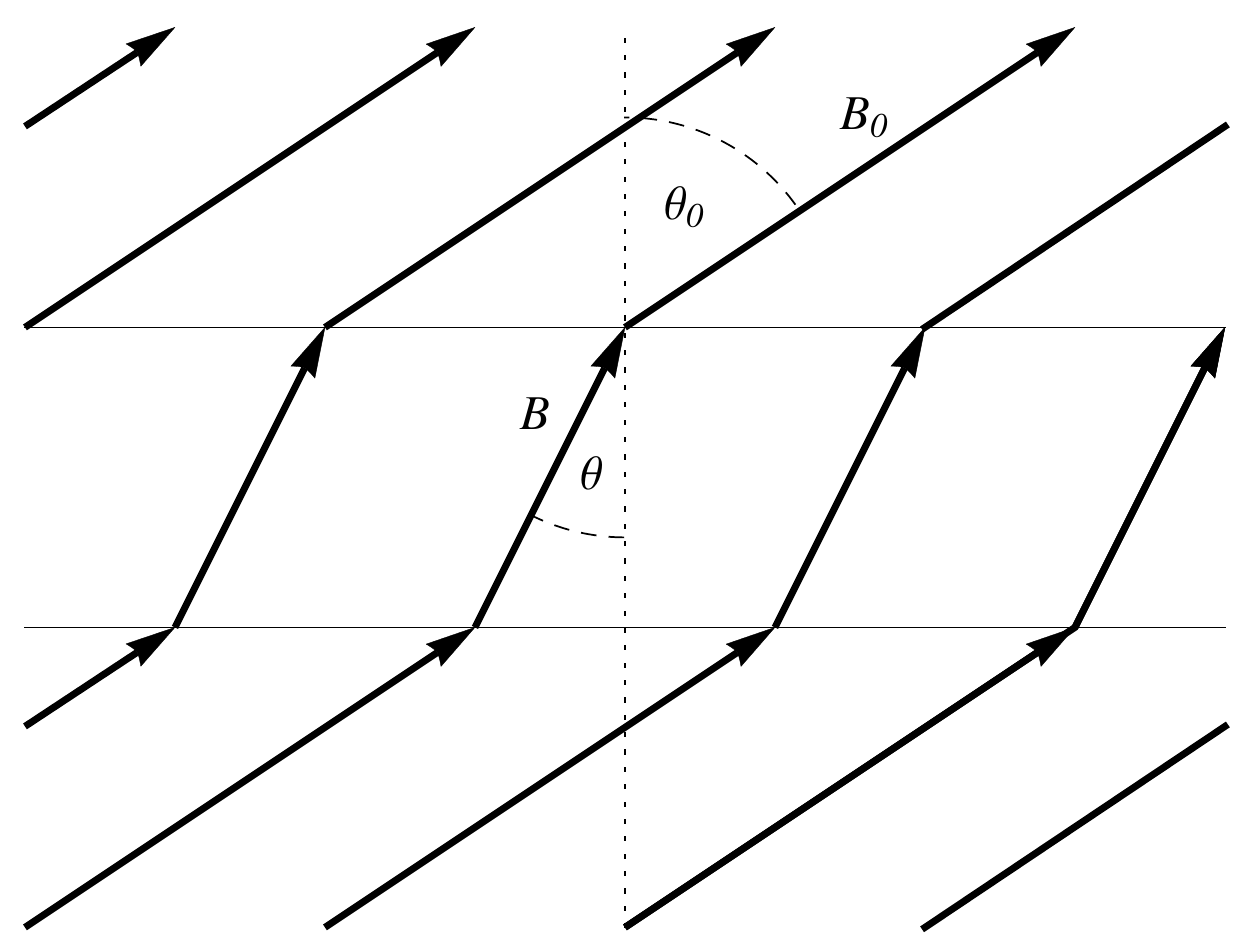}
\caption{%
Superconducting slab subjected to an applied magnetic induction $\bm B_0$ of magnitude $B_0=\mu_0 H_0  = f_0 B_c$ and angle $\theta_0$.  The average magnetic induction in the superconductor $\bm B$ has magnitude $B = f_n B_c$ and angle $\theta$.}
\label{BFig}
\end{figure}

When a magnetic field $\bm H_0$ (magnetic induction $\bm B_0 = \mu_0 \bm H_0$) with components $H_{0z} = H_0 \cos \theta_0$ and $H_{0x} = H_0 \sin \theta_0$  (See Fig.\ \ref{BFig} for a definition of $\theta$ and $\theta_0$) produces the intermediate state in an ellipsoid of revolution, the internal fields $\bm B = \mu_0 (\bm H + \bm M)$ are all parallel to $\hat B = \hat z B \cos \theta + \hat x B \sin \theta$, and their magnitudes are related via $B = f_n B_c = \mu_0 H_c(1+\chi).$  With $f_0 = H_0/H_c$, the demagnetization boundary-condition equations connecting $f_0$, $\theta_0$, $f_n$, and $\theta$ are \cite{Cape67,Goldfarb92}
\begin{eqnarray}
\cos\theta= \frac{f_0 \cos \theta_0}{1-N_z+N_z f_n}, \label{cosNz}\\
\sin\theta= \frac{f_0 \sin \theta_0}{1-N_x+N_x f_n},  \label{sinNx}
\end{eqnarray} 
where $2N_x + N_z = 1$ by the demagnetization coefficient sum rule.  These equations can be solved numerically to determine  $f_n$ and $\theta$ as functions of $f_0$ and  $\theta_0$

We turn now to the case of flat samples of finite thickness $d$ but infinite lateral dimensions, as shown in Fig.\ \ref{BFig}.   According to Maxwell's equations, the continuity of the perpendicular component of $\bm B$ and the tangential component of $\bm H$ requires that
\begin{eqnarray}
B \cos \theta &=& B_0 \cos \theta_0, \label{Bnorm}\\
H_c \sin \theta &=& H_0 \sin \theta_0. \label{Htan}
\end{eqnarray}
Note that these equations are equivalent to Eqs.\ (\ref{cosNz}) and (\ref{sinNx}) in the limit as $\gamma \to 0$, $N_z \to 1$, and $N_x \to 0$.

\begin{figure}
\includegraphics[width=8cm]{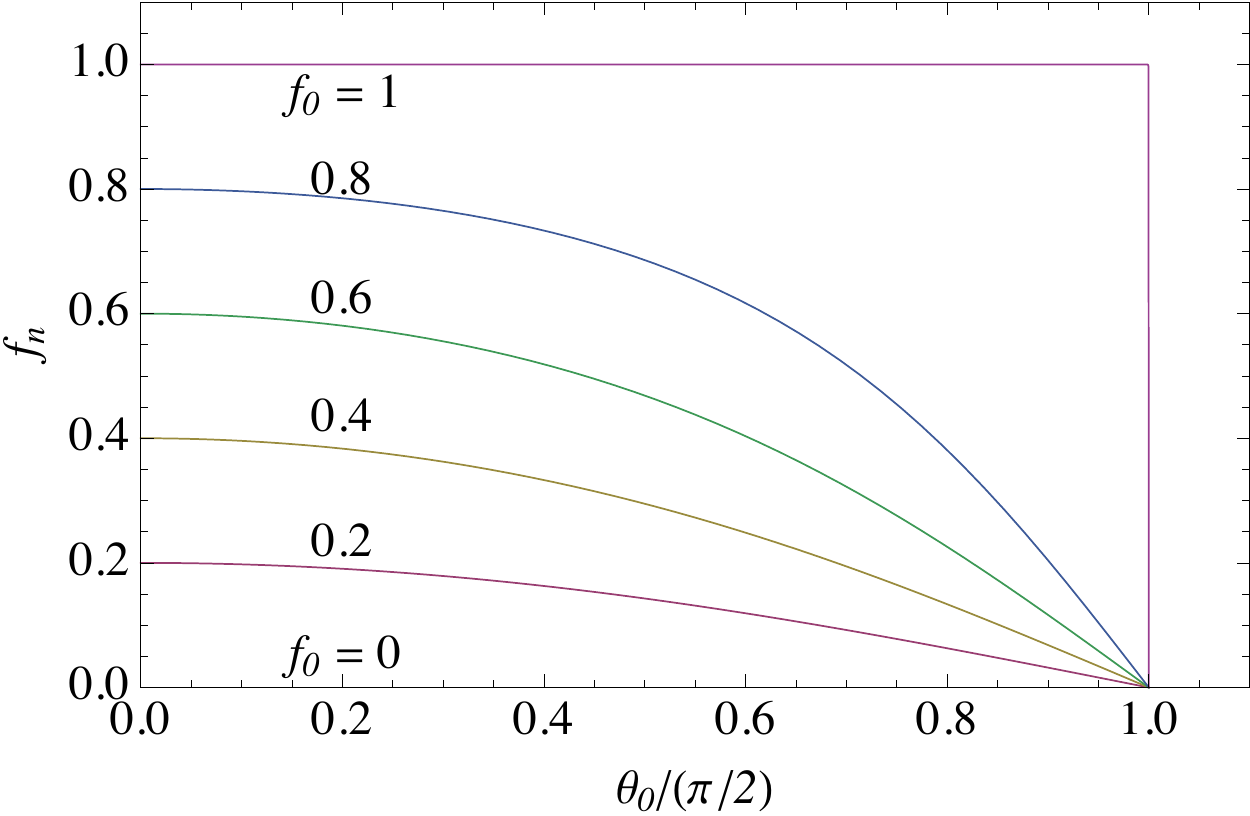}
\caption{%
Normal fraction $f_n = B/B_c$ in the superconducting slab vs applied field angle $\theta_0$ for various values of $f_0 = B_0/B_c$.}
\label{fnFig}
\end{figure}

\begin{figure}
\includegraphics[width=8cm]{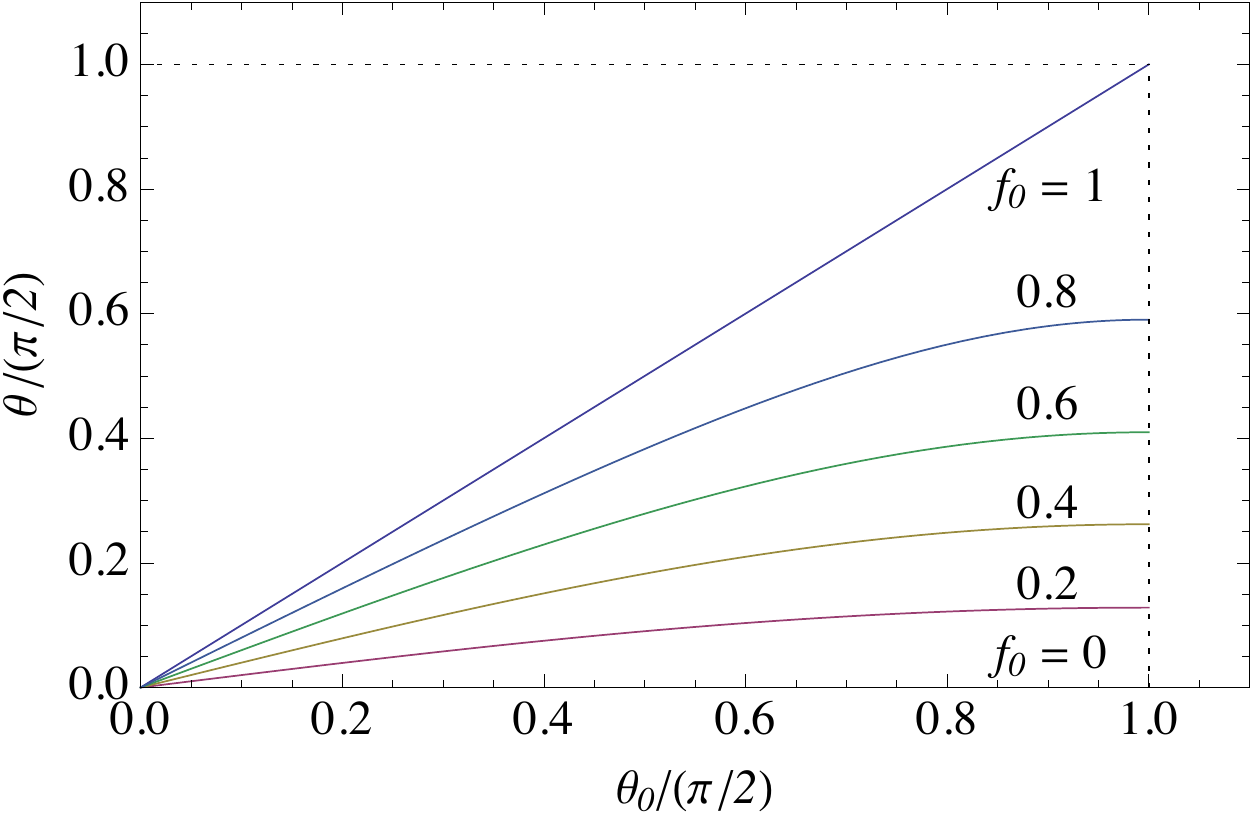}
\caption{%
Angle $\theta$ of the average magnetic induction $\bm B$ in the superconducting slab vs applied field angle $\theta_0$ for various values of $f_0 = B_0/B_c$. }
\label{thetaFig}
\end{figure}

It is generally not possible to look inside the superconductor to determine $f_n$ and $\theta$, but expressions for these quantities in terms of $f_0 = B_0/B_c= H_0/H_c$
and $\theta_0$ can be obtained from Eqs.\ (\ref{Bnorm}) and (\ref{Htan}):
\begin{eqnarray}
f_n &=& \frac{ f_0 \cos \theta_0}{\sqrt{1-f_0^2 \sin ^2 \theta_0}}, \label{fn}\\
\theta &=& \sin^{-1}(f_0 \sin \theta_0). \label{theta}
\end{eqnarray}
Figures \ref{fnFig} and \ref{thetaFig} show the behavior of $f_n$ and $\theta$ as calculated from Eqs.\ (\ref{fn}) and (\ref{theta}).  Note that  $f_n \to 1$ and $\theta \to \theta_0$ as $f_0 \to 1$, but that $f \to f_0 \cos \theta_0$ and $\theta \to 0$ as $f_0 \to 0$.  The case $\theta_0= \pi/2$ is singular, because in this case the applied field is exactly parallel to the surface of the infinite slab and there is no intermediate state.   $B$ remains zero inside the sample until $H_0$ reaches $H_c$, at which point superconductivity in the bulk is quenched, and $B$ jumps to $B_c$.  

When the intermediate state consists of an array of flux tubes of radius $R$, each carrying  magnetic flux $\Phi = \pi R^2 B_c$, Eq.\  (\ref{Htan}) can be interpreted as a force-balance equation.  The line tension (energy per unit length of flux tube), accounting for condensation and magnetic-field energy costs, is $T_\ell = H_c \Phi.$   The horizontal component of the applied field $H_0 \sin\theta_0$ (see Fig.\ \ref{BFig})  generates a sheet-current density of magnitude $K_{\parallel} = H_0 \sin\theta_0$ on the top and bottom surfaces.  The corresponding Lorentz forces of magnitude $F_\parallel = K_{\parallel}\Phi$ pull the top of the flux tube to the right and the bottom of the flux tube to the left.  The forces are balanced when $F_\parallel$ is equal to the horizontal component of the line tension, i.e., when $ H_0 \Phi\sin\theta_0  = H_c\Phi\sin\theta$,  which is equivalent to Eq.\ (\ref{Htan}).  However, in the static case the local sheet current and magnetic field must redistribute around the ends of the flux tube, so that the induced current remains a supercurrent and avoids being driven through the normal domain.

\section{Models for the magnetic structure in the intermediate state \label{IS}}

\subsection{Competing contributions to the free energy\label{Competing}}


When a perpendicular magnetic induction $B_0$ is applied to a flat type-I superconductor ($\theta_0 = 0$ in Fig.\ \ref{BFig}), we have $\theta = 0$ [Eq.\ (\ref{theta})], the average magnetic induction $B$ inside the superconductor becomes equal to $B_0$ [see Eq.\ (\ref{Bnorm})], and the normal fraction averaged over the sample volume is $f_n = B/B_c = f_0 = B_0/B_c$.   On the other hand,  the local magnetic induction $b$ has an extremely complicated spatial dependence.\cite{Landau1938,Landau1943,Shoenberg1952,Livingston1969,Tinkham1996,Poole2007,Huebener2001}

As revealed by magneto-optical observations, the magnetic structure for $f_n \ll 1$ can be described crudely as an array of isolated normal domains or flux tubes carrying magnetic flux $\Phi = N \phi_0$, i.e., containing an integer number $N$ of superconducting flux quanta $\phi_0 = h/2e$, surrounded by superconducting regions.  The radius $r$ of such a domain is given by $\Phi = B_c \pi r^2$.  As $B_0$ increases, fluxoid quantization keeps the flux tubes the same size,  and to allow $B$ to increase, additional flux tubes must move into the sample from the edges.

As $f_n$ increases, the normal domains or flux tubes tend to connect and form stripe-like segments, and for $f_n = 1/2$, the complex magnetic structure can be described roughly as arrays of alternating normal and superconducting stripe-like domains of roughly equal area.

As  $f_n \to 1$, the magnetic structure can be described as an array of isolated superconducting domains surrounded by normal material.  These domains may occur as stripe-like segments or as nearly round spots. The magneto-optical images resemble  mirror images of those for $f_n \ll 1$.  What is very different, however, is that fluxoid quantization places no constraints on the sizes of the superconducting domains, and an increase of $B_0$ tends to make the superconducting domains shrink as the surrounding normal material grows in total area.

The basic structure of the intermediate state in disks, plates, slabs, foils, and single crystals of type-I superconductors is known to be determined chiefly by two competing energy contributions:\cite{Landau1938,Landau1943,Shoenberg1952,Livingston1969,Tinkham1996,Poole2007,Huebener2001} (a) a positive wall energy  between a  normal domain containing a local flux density $b = \mu_0 H_c$ and a Meissner domain of zero flux density and (b) the excess energy of the nonuniform magnetic field outside the surface relative to that of a uniform magnetic-field distribution.  The wall-energy cost favors well-separated  large normal domains, while the magnetic-energy cost favors finely divided normal domains.
The wall energy $\gamma$ per unit of surface area is given in SI units by\cite{Tinkham1996}
\begin{equation}
\gamma=(B_c^2/2\mu_0)\delta,\label{gamma}
\end{equation}
where the wall-energy parameter\cite{Huebener2001} $\delta$ has units of length.
Minimization of the total energy cost in a sample of thickness $d$ leads to a domain structure characterized by a length scale proportional to $(\delta d)^{1/2}$.

For small $d$, the length scale of the domain structure and the normal-domain size, which also scales as $(\delta d)^{1/2}$,  become very small.  Accordingly, the number of flux quanta $N$ in a flux tube also becomes very small, and for sufficiently small $d$, it is found that $N = 1$, so that the magnetic structure becomes equivalent to that in type-II superconductors.  In this paper we consider the opposite limit, for which $N \gg 1$.

As noted by Tinkham, \cite{Tinkham1996} calculations assuming different magnetic structures including the competing wall-energy and field-energy contributions lead to expressions for the free energy that differ numerically only slightly.  This helps to  explain why magneto-optical observations upon field  cycling  show somewhat different magnetic structures each time a sample is exposed to the same $B_0$ at the same temperature $T$.  It is also important to note that there are free-energy barriers that prevent the sample from assuming the magnetic structure corresponding to the global free-energy minimum for a given $B_0$ and $T$.  As $B_0$ and $T$ change, it is likely that the sample gets stuck in a spatial configuration with a local free-energy minimum not far from the global minimum.  These effects inevitably lead to significant history effects; the appearance of the magnetic structure for a given $B_0$ and $T$ depends strongly upon the sample's field and temperature history.

\subsection{Modeling the magnetic structure in a perpendicular field\label{PerpFieldSec}}

We present here three models of the magnetic structure in a homogeneous, isotropic type-I superconductor of thickness $d$.   For small values of $f_n
 = B/B_c$ (case 1) we approximate the magnetic structure as an equilateral triangular array (with lattice parameter $D$) of normal cylindrical flux spots of radius  $R$ and magnetic flux density $B_c$ surrounded by the superconducting phase.   Similarly, for large values of $f_n$ (case 3) we approximate the magnetic structure as an equilateral triangular array (with lattice parameter $D$) of superconducting cylinders of radius $R$ surrounded by normal material of flux density $B_c$. For intermediate values of   $f_n$ (case 2) we approximate the magnetic structure as a periodic array (with period $D$) of parallel normal domains of width $W$ and magnetic flux density $B_c$ separated by the superconducting phase.  We evaluate the free energy per unit sample volume using the same method for all three cases and identify the best model for a given $f_n$ as that with the smallest free energy.  For each case, we calculate the free energy per unit sample volume as  the sum of wall-energy and field-energy contributions.
To find the equilibrium topology, we need to consider only those contributions to the free energy that differ for the different models; e.g., we omit the superconducting condensation energy from the calculations.
Here we follow the commonly used convention of expressing the normal fraction in a perpendicular field as $h = f_n = B/B_c = f_0 = B_0/B_c$.

\subsubsection{Case 1, small $h$\label{Case1}}

Here we deal with normal cylindrical flux spots of radius  $R$ surrounded by  superconducting phase. Within the unit cell of volume $\sqrt{3}D^2d/2$, the area of the normal-superconducting interface is $2\pi R d$.  From Eq.\ (\ref{gamma}), we see that   the wall-energy cost of the intermediate state per unit sample volume is
\begin{equation}
F_1 = \frac{2\pi B_c^2 \delta R}{\sqrt{3} \mu_0 D^2}=\frac{B_c^2 \delta R}{\mu_0 R_0^2}. \label{F11}
\end{equation}
Here the Wigner-Seitz radius  $R_0 = (\sqrt{3}/2\pi)^{1/2}D$ is chosen such 
that
the circular cross-sectional area is the same as the unit-cell cross-sectional area.

The field-energy cost of the intermediate state per unit sample volume is
\begin{equation}
F_2 = \frac{1}{\pi \mu_0 R_0^2 d} \int dV(b^2-B^2), \label{F21int}
\end{equation}
where the integral is to be carried out within the Wigner-Seitz cylinder above the sample surface at $z = 0$.  Here $\bm b =\hat \rho b_\rho(\rho,z) + \hat z b_z(\rho,z)= -\nabla \phi$ is the magnetic induction within the cylinder, subject to the boundary conditions $b_z(\rho,0) = B_c$ for $\rho < R$, $b_z(\rho,0) = 0$ for $R < \rho < R_0$, $b_z(\rho,\infty) = B,$ and  $b_\rho(R_0,z) = 0$.  This is a readily solvable boundary-value problem in cylindrical coordinates,\cite{Jackson62}   and after application of the divergence theorem, we find that the integral in Eq.\ (\ref{F21int}) is proportional to $S_1(R/R_0)$, where
\begin{equation}
S_1(u) = \sum_{n=1}^\infty \frac{1}{x_{1n}^3}\Big[\frac{J_1(x_{1n}u)}{J_0(x_{1n})}\Big]^2, \label{S}
\end{equation}
where $J_m(x)$ is the Bessel function of order $m$, and $x_{1n}$ is the $n$-th root of $J_1(x)$ (e.g., $x_{11}$ = 3.83, $x_{12}$ = 7.02, $x_{13}$ = 10.17, etc.).  Our method for calculating the field energy in cases 1, 2, and 3 is similar to that used in the current-loop model by Goldstein et al.\cite{Goldstein1996} 

Taking the sum of $F_1$ and $F_2$ and making use of $R/R_0 = \sqrt{h}$, we find that the net cost in free energy per unit sample volume for case 1 is
\begin{equation}
\Delta F_1 = \frac{B_c^2}{\mu_0}\Big[\frac{\delta \sqrt{h}}{R_0}+\frac{4 R_0 h S_1(\sqrt{h})}{d}\Big].
\end{equation}
For a given $h$, the wall-energy term favors large length scales $R_0$, while the field-energy term favors small length scales.  At the value of $R_0$ that minimizes $\Delta F_1$, we obtain
\begin{eqnarray}
R_0 &=& \frac{\sqrt{\delta dh}}{2\Phi_1}, \label{R0min1} \\
R &=& \frac{h\sqrt{\delta d}}{2\Phi_1}, \label{Rmin1} \\
\Delta F_1 &=& \frac{4B_c^2}{\mu_0}\Big(\frac{\delta}{d}\Big)^{1/2}\Phi_1, \label{DeltaF1}\\
\Phi_1 &=& [h^{3/2} S_1(\sqrt{h})]^{1/2}.\label{Phi1}
\end{eqnarray}
See the dashed curve in Fig.\ \ref{Phi123Fig}.

\subsubsection{Case 2, moderate $h$\label{Case2}}

Next we consider flux-filled normal domains of width $W = 2R$ parallel to the $y$ axis with periodicity $D = 2R_0$ along the $x$ direction.  The normal domains alternate with flux-free superconducting domains.  The area of normal-superconducting interface per unit sample volume  is $2/D$, and from Eq.\ (\ref{gamma}), we see that   the wall-energy cost of the intermediate state per unit sample volume is
\begin{equation}
F_1 = \frac{B_c^2 \delta}{\mu_0 D} = \frac{B_c^2 \delta}{2\mu_0 R_0}. \label{F12}
\end{equation}

The field-energy cost of the intermediate state per unit sample volume is
\begin{equation}
F_2 = \frac{1}{2 \mu_0 R_0 d} \int dA(b^2-B^2), \label{F22int}
\end{equation}
where the integral is to be carried out over the area of width $D = 2R_0$ and infinite height above the sample surface at $z = 0$.   Here $\bm b =\hat x b_x(x,z) + \hat z b_z(x,z)= -\nabla \phi$ is the magnetic induction within this area, subject to the boundary conditions $b_z(x,0) = B_c$ for $|x| < R$, $b_z(x,0) = 0$ for $R < |x| < R_0$, $b_z(x,\infty) = B,$ and  $b_x(\pm R_0,z) = 0$.  This is another readily solvable boundary-value problem,   and after applying the divergence theorem and  making use of $R/R_0 = W/D = h$, we find that the integral in Eq.\ (\ref{F22int}) is proportional to $S_2(h)$,
where
\begin{equation}
S_2(h) = \sum_{n=1}^\infty \frac{\sin^2(n\pi h)}{(n\pi)^3}, \label{S2}
\end{equation}
which can be expressed in terms of the Riemann zeta function $\zeta(n)$ and the polylogarithm function $Li_n(z)$ as
\begin{equation}
S_2(h) = [2\zeta(3)-Li_3(e^{i2\pi h})-Li_3(e^{-i2\pi h})]/4\pi^3. \label{S2polylog}
\end{equation}
It is easily shown from Eq.\ (\ref{S2}) that $S_2(h)$ is symmetric about $h = 1/2$; $S_2(1-h) = S_2(h)$.

Taking the sum of $F_1$ and $F_2$, we find that the net cost in free energy per unit sample volume for case 2 is
\begin{equation}
\Delta F_2 = \frac{B_c^2}{\mu_0}\Big[\frac{\delta }{2 R_0}+\frac{2 R_0 S_2(h)}{d}\Big].
\end{equation}
For a given $h$, the wall-energy term again favors large length scales $R_0$, while the field-energy term favors small length scales.  At the value of $R_0$ that minimizes $\Delta F_2$, we obtain
\begin{eqnarray}
R_0 &=& \frac{\sqrt{\delta d}}{4\Phi_2}, \label{R0min2} \\
R &=& \frac{h\sqrt{\delta d}}{4\Phi_2}, \label{Rmin2} \\
\Delta F_2 &=& \frac{4B_c^2}{\mu_0}\Big(\frac{\delta}{d}\Big)^{1/2}\Phi_2, \label{DeltaF2}\\
\Phi_2 &=& [S_2(h)]^{1/2}/2.\label{Phi2}
\end{eqnarray}
See the solid curve in Fig.\ \ref{Phi123Fig}, and note the mirror symmetry $\Phi_2(h)=\Phi_2(1-h)$.

\subsubsection{Case 3, large $h$\label{Case3}}

Here we consider cylindrical flux-free superconducting regions of radius  $R$ surrounded by  flux-filled normal phase.  Within the unit cell of volume $\sqrt{3}D^2d/2$, the area of the normal-superconducting interface is $2\pi R d$, and the wall-energy cost of the intermediate state per unit sample volume is the same as in case 1,
\begin{equation}
F_1 = \frac{2\pi B_c^2 \delta R}{\sqrt{3} \mu_0 D^2}=\frac{B_c^2 \delta R}{\mu_0 R_0^2}. \label{F13}
\end{equation}
As in Sec.\ \ref{Case1}, the Wigner-Seitz radius  $R_0 = (\sqrt{3}/2\pi)^{1/2}D$ is again chosen such that the circular cross-sectional area is the same as the unit-cell cross-sectional area.

The field-energy cost of the intermediate state per unit sample volume is
\begin{equation}
F_2 = \frac{1}{\pi \mu_0 R_0^2 d} \int dV(b^2-B^2), \label{F23int}
\end{equation}
where the integral is to be carried out within the Wigner-Seitz cylinder above the surface of the sample.  Here $\bm b =\hat \rho b_\rho(\rho,z) + \hat z b_z(\rho,z)= -\nabla \phi$ is the magnetic induction within the cylinder, subject to the boundary conditions $b_z(\rho,0) = 0$ for $\rho < R$, $b_z(\rho,0) = B_c$ for $R < \rho < R_0$, $b_z(\rho,\infty) = B,$ and  $b_\rho(R_0,z) = 0$.  This again is a readily solvable boundary-value problem in cylindrical coordinates,   and after application of the divergence theorem, we find that the integral in Eq.\ (\ref{F21int}) is  proportional to $S_1(R/R_0)$.

Taking the sum of $F_1$ and $F_2$ and making use of $R/R_0 = \sqrt{f_s}$, where $f_s = 1-h$, we find that net cost in free energy per unit sample volume for case 3 is
\begin{equation}
\Delta F_3 = \frac{B_c^2}{\mu_0}\Big[\frac{\delta \sqrt{f_s}}{R_0}+\frac{4 R_0 f_s S_1(\sqrt{f_s})}{d}\Big].
\end{equation}
For a given $f_s$, the wall-energy term again favors large length scales $R_0$, while the field-energy term favors small length scales.  At the value of $R_0$ that minimizes $\Delta F_3$, we obtain
\begin{eqnarray}
R_0 &=& \frac{\sqrt{\delta d(1-h)}}{2\Phi_3}, \label{R0min3} \\
R &=& \frac{h\sqrt{\delta d}}{2\Phi_3}, \label{Rmin3} \\
\Delta F_1 &=& \frac{4B_c^2}{\mu_0}\Big(\frac{\delta}{d}\Big)^{1/2}\Phi_3, \label{DeltaF3}\\
\Phi_3(h) &=& \Phi_1(1\!-\!h) =[(1\!-\!h)^{3/2} S_1(\sqrt{1\!-\!h})]^{1/2}.\label{Phi3}
\end{eqnarray}
See the dot-dashed curve in Fig.\ \ref{Phi123Fig}, and note that $\Phi_3(h)$, which describes a triangular array of superconducting cylinders, is the mirror image of $\Phi_1(h)$, which describes a triangular array of normal cylinders.

\subsubsection{Lowest-free-energy models, arbitrary $h$\label{Phi0Sec}}

\begin{figure}
\includegraphics[width=8cm]{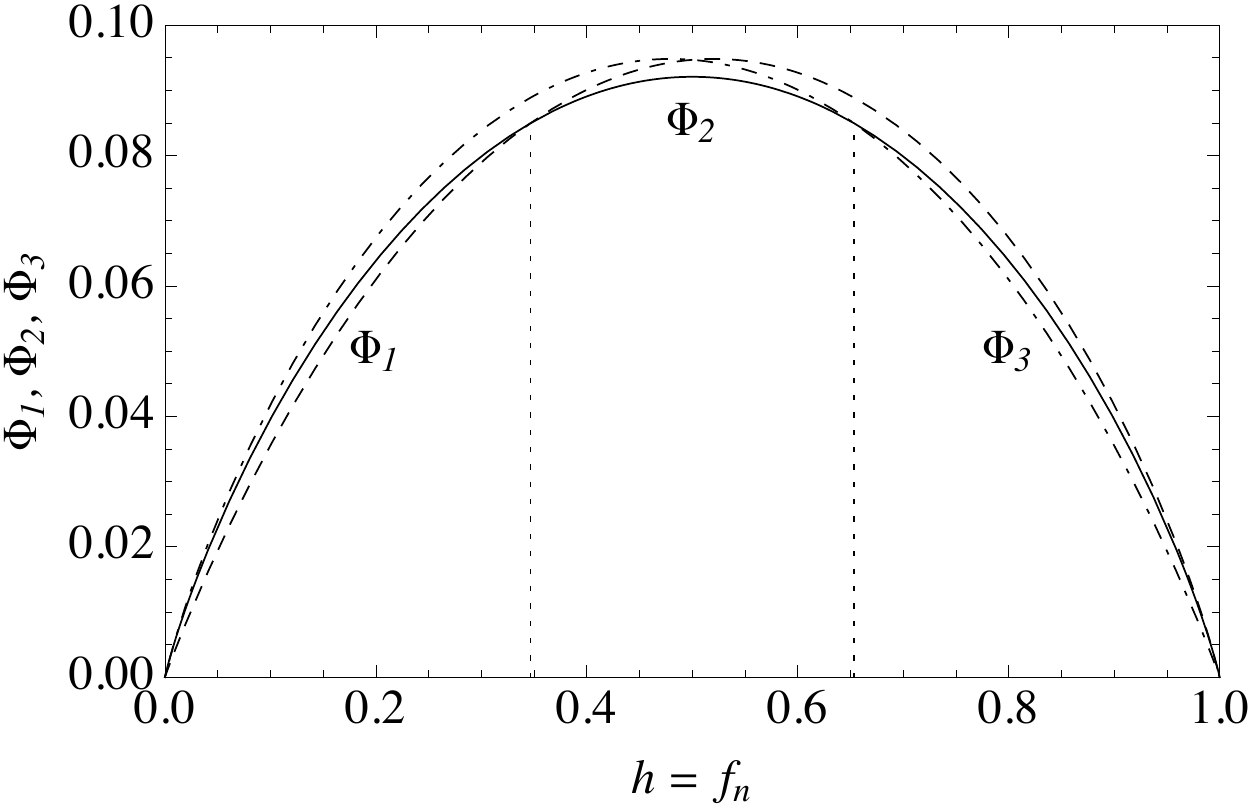}
\caption{%
Plots of the normalized free energies for case 1: $\Phi_1$ (dashed), triangular array of cylindrical normal flux tubes, Eq.\ (\ref{Phi1}); case 2: $\Phi_2$ (solid), parallel array of normal and superconducting domains, Eq.\ (\ref{Phi2}); and  case 3: $\Phi_3$ (dot-dashed), triangular array of superconducting cylinders, Eq.\ (\ref{Phi3}).  The free energy cost $\Delta F$ is smallest for case 1 when $0 \le h \le 0.346$, case 2 for $0.346 \le h \le 0.654$, and case 3 for $0.654 \le h \le 1$.}
\label{Phi123Fig}
\end{figure}

The dependencies of $\Phi_1$, $\Phi_2$, and $\Phi_3$ upon $h$ are shown  in Fig.\ \ref{Phi123Fig}.  As expected, $\Phi_1$ is favored for small $h$, $\Phi_2$ for intermediate $h$, and $\Phi_3$ for large $h$.  Within the restrictive assumptions of this paper (constant flux density $B_c$ in straight normal domains), our best model for the intermediate state is $\Phi_0(h)$, defined to be equal to $\Phi_1(h)$ when $0 \le h \le 0.346$,  $\Phi_2(h)$ when $0.346 \le h \le 0.654$, and $\Phi_3(h)$  when $0.654 \le h \le 1$. This is shown as the solid curve in Fig.\ \ref{PhiCompsFig}.

\subsubsection{Comparison with other models}

\begin{figure}
\includegraphics[width=8cm]{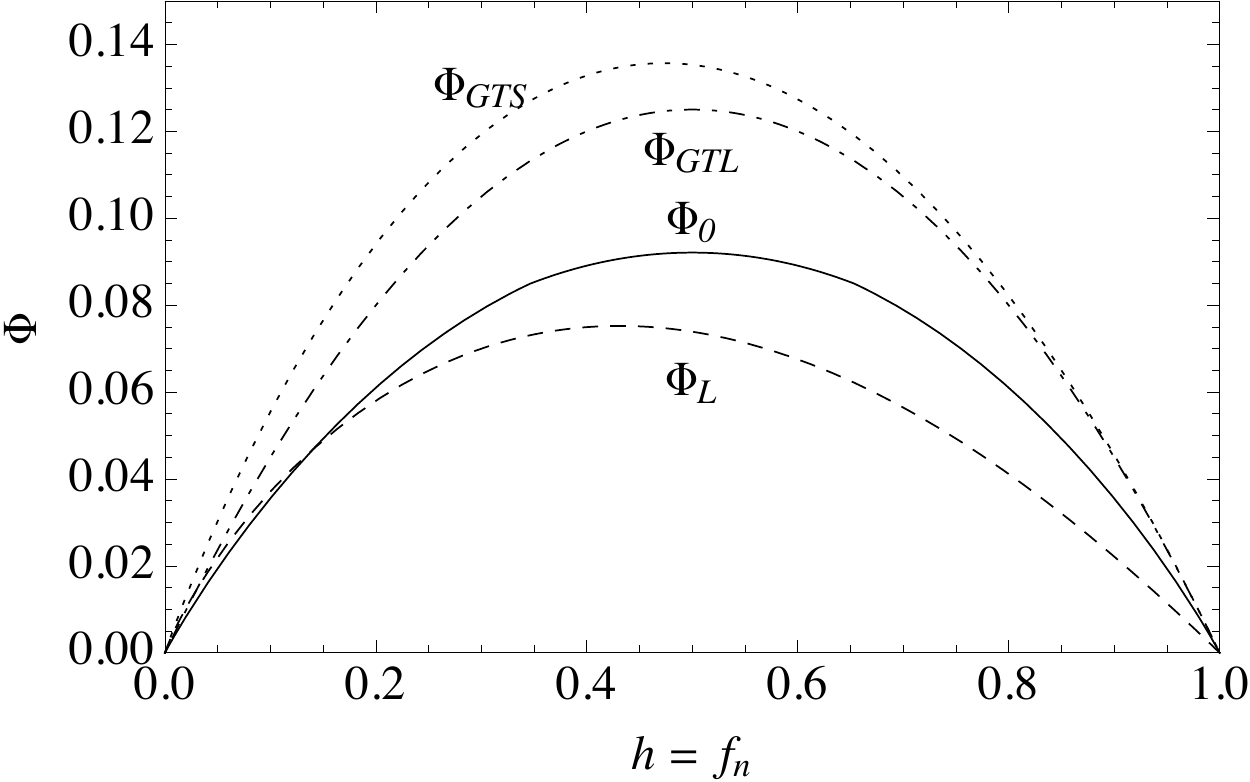}
\caption{%
Plots of the normalized free energies for our model $\Phi_0$ (solid, defined in Sec.\ \ref{Phi0Sec}), the Landau laminar-domain model $\Phi_L$ (dashed, Eq.\ (\ref{PhiLandau})), the Goren-Tinkham laminar-domain model $\Phi_{GTL}$ (dot-dashed, Eq.\ (\ref{PhiGTL})), and the  the Goren-Tinkham flux-spot model $\Phi_{GTS}$ (dotted, Eq.\ (\ref{PhiGTS}))
}
\label{PhiCompsFig}
\end{figure}

Landau's classic calculation of a laminar domain structure,\cite{Landau37,Fortini1972,Landau84} which accounted for bending of the normal-superconducting interfaces, yielded a function $f(h)$.  For a parallel domain structure of periodicity length $D$,  normal domain width $W$ (deep inside the sample), and free-energy  cost per unit sample volume $\Delta F_L$, these quantities are given by
\begin{eqnarray}
D &=& \frac{\sqrt{\delta d}}{2\Phi_L}, \label{Dmin} \\
W &=& \frac{h\sqrt{\delta d}}{2\Phi_L}, \label{Wmin} \\
\Delta F_L &=& \frac{4B_c^2}{\mu_0}\Big(\frac{\delta}{d}\Big)^{1/2}\Phi_L, \label{DeltaFLandau}\\
\Phi_L &=& [f(h)]^{1/2}/2.\label{PhiLandau}
\end{eqnarray}
See the dashed curve in Fig.\ \ref{PhiCompsFig}.  It is notable that $\Phi_L(h)$ does {\it not} have mirror symmetry about $h = 1/2$.  The reason for this is that the Landau calculation does not deal with straight tubes and lamellae but instead includes additional contributions to the free energy associated with normal domains that bend outwards and superconducting domains that bend inwards as they approach the surface.

Note from Fig.\ \ref{PhiCompsFig} that the laminar-domain Landau  curve $\Phi_L(h)$ (dashed)  lies below the  curve for $\Phi_0$ (solid) for moderate and larger values of $h$.  Evidently this indicates that, by neglecting the effects of bending the normal-superconducting interfaces, our model for $\Phi_0$ overestimates the free-energy cost of the intermediate state and underestimates the length scales ($D$, $R_0$, $W$, and $R$) of the actual magnetic structure.  On the other hand, the  Landau  curve $\Phi_L(h)$ (dashed)  lies {\it above} the  curve for $\Phi_0$ (solid) for small values of $h$.  This evidently indicates that a laminar-domain model cannot be applied  to model accurately the magnetic structure for small $h$, which is better described as an array of isolated normal domains.

Goren and Tinkham\cite{Goren71} proposed two models for the intermediate state based on the assumption of straight normal domains containing constant flux density $B_c$ but estimating the field-energy contributions to the free energy using a healing-length approximation.\cite{Tinkham1996}  For a parallel domain structure of periodicity length $D$,  normal domain width $W$, and free-energy  cost per unit sample volume $\Delta F_{GTL}$, these quantities are given by\cite{Goren71}
\begin{eqnarray}
D &=& \frac{\sqrt{\delta d}}{2\Phi_{GTL}}, \label{DminGTL} \\
W &=& \frac{h\sqrt{\delta d}}{2\Phi_{GTL}}, \label{WminGTL} \\
\Delta F_{GTL} &=& \frac{4B_c^2}{\mu_0}\Big(\frac{\delta}{d}\Big)^{1/2}\Phi_{GTL}, \label{DeltaFGTL}\\
\Phi_{GTL} &=& h(1-h)/2.\label{PhiGTL}
\end{eqnarray}
See the dot-dashed curve in Fig.\ \ref{PhiCompsFig}.

The second model considered by Goren and Tinkham\cite{Goren71} consisted of an equilateral triangular array of flux spots. For an equilateral triangular array (lattice parameter $D$) of hexagonal normal domains of width $W$, and free-energy  cost per unit sample volume $\Delta F_{GTS}$, these quantities are given by\cite{Goren71}
\begin{eqnarray}
D &=& \frac{\sqrt{\delta dh}}{\Phi_{GTS}}, \label{DminGTS} \\
W &=& \frac{h\sqrt{\delta d}}{\Phi_{GTS}}, \label{WminGTS} \\
\Delta F_L &=& \frac{4B_c^2}{\mu_0}\Big(\frac{\delta}{d}\Big)^{1/2}\Phi_{GTS}, \label{DeltaFGTS}\\
\Phi_{GTS} &=& [h^2(1-h)(1-h^{1/2})/2.\label{PhiGTS}
\end{eqnarray}
See the dotted curve in Fig.\ \ref{PhiCompsFig}.

The main difference between the Goren-Tinkham\cite{Goren71} approach and that used for  $\Phi_1$, $\Phi_2$, $\Phi_3$, and $\Phi_0$ is the method used to calculate the excess energy of the nonuniform magnetic field outside the surface relative to that of a uniform magnetic-field distribution.  Our calculations show that the healing-length approximation\cite{Goren71,Tinkham1996} overestimates the field-energy contribution $F_2$ and underestimates the length scales ($D$ and $W$) of the actual magnetic structure.  The Goren-Tinkham models\cite{Goren71} have the property that $\Phi_{GTL} < \Phi_{GTS}$, predicting that the lamellar structure is favored for all $h$.  This finding is a result of the fact that the authors used different healing-length approximations for the lamellar and spot structures.
However, to be consistent with experimental observations, models of the intermediate state should have  the smallest free energy costs for arrays of separated normal domains for small $h$, stripe-like or parallel domains for $h \sim 1/2$, and arrays of separated superconducting domains for large $h$.

\subsection{Modeling the magnetic structure when both parallel and perpendicular fields are applied\label{InclinedSec}}

It is well known that application of a parallel magnetic field to a sample containing magnetic structure produced by a perpendicular field tends to orient the magnetic structure.\cite{Sharvin57,Dzyaloshinskii55}  Here we use the approach of Sec.\ \ref{PerpFieldSec} to consider the conditions for which the Sharvin laminar structure, with laminar domains oriented along the parallel component of the applied field, is energetically favorable.  We consider, for an applied magnetic field of magnitude $H_0$ and angle $\theta_0$, cases 1$^\prime$, 2$^\prime$, and 3$^\prime$, which are tilted-field extensions of cases 1, 2, and 3.
We evaluate the free energy per unit sample volume using the same method for all three cases and identify the best model for a given $f_n$ as that with the smallest free energy.  For each case, we calculate the free energy per unit sample volume as  the sum of wall-energy and field-energy contributions.
These calculations differ from those of  Sec.\ \ref{PerpFieldSec} in part because the normal fraction $f_n$ must be determined from Eq.\ (\ref{fn}).

\subsubsection{Case 1$^{\prime}$, small $f_n$\label{Case1'}}

We begin by considering the magnetic structure inside the superconductor for small values of $f_n  = B/B_c$ (case 1$^{\prime}$, as an 
extension of case 1).  We approximate the magnetic structure as an array of quantized identical cylindrical flux tubes, oriented parallel to $\hat B$ at angle $\theta$, as shown in Fig.\ \ref{BFig}.  In the plane normal to $\hat B$ the flux tubes are assumed to arrange themselves as an equilateral triangular array (with lattice parameter $D$) of circular normal  flux tubes of radius  $R$ and core magnetic-flux density $B_c$  surrounded by the superconducting phase.  For doing calculations of the magnetic energy of the structure above and below the sample, however, it is convenient to replace the equilateral triangular array inside the sample with a Wigner-Seitz cylinder of radius  $R_0 = (\sqrt{3}/2\pi)^{1/2}D$ with the same  area as the unit cell.
This Wigner-Seitz cylinder is parallel to $\hat B$.
Within the Wigner-Seitz cylinder of  volume $\pi R_0^2d/\cos\theta$, the area of the normal-superconducting interface is $2\pi R d/\cos\theta$.  From Eq.\ (\ref{gamma}), we see that   the wall-energy cost of the intermediate state per unit sample volume is
\begin{equation}
F_1 =\frac{B_c^2 \delta R}{\mu_0 R_0^2}. \label{F11'}
\end{equation}

The Wigner-Seitz cylinder intersects the top surface at the angle $\theta$, so that the intersection is an ellipse of semimajor axis $R_{0s} = R_0/\cos\theta$ along the $x$ direction and semiminor axis $R_0$ along the $y$ direction.
Similarly, the intersection of the flux tube with the surface is  an ellipse of semimajor axis $R_{s} = R/\cos\theta$ along the $x$ direction and semiminor axis $R$ along the $y$ direction.  
The $z$ component of the magnetic flux density emerging from the latter area is $b_z = B_c  \cos\theta$, but the area is $\pi R^2/\cos\theta$, so that the magnetic flux emerging is $\Phi = \pi R^2 B_c$, the same as the magnetic flux carried by a flux tube inside the sample.

The field-energy cost of the intermediate state per unit sample volume is
\begin{equation}
F_2 = \frac{\cos\theta}{\pi \mu_0 R_0^2 d} \int dV(b^2-B^2\cos^2\theta), \label{F21'int}
\end{equation}
where the integral is to be carried out within a new Wigner-Seitz elliptical cylinder above the sample surface at $z = 0$. However, we can make use of a scale transformation similar to that used in Ref.\ \onlinecite{Klemm80} to reexpress the integral in terms of new primed coordinates.  The field-energy cost of the intermediate state per unit sample volume becomes

\begin{equation}
F_2 = \frac{\cos\theta}{\pi \mu_0 R_0^2 d} \int dV'(b'^2-B^2\cos^2\theta), \label{F21''int}
\end{equation}
where the integral is to be carried out within a new Wigner-Seitz cylinder of radius $R_0/\sqrt{\cos\theta}$ above the sample surface at $z' = 0$, and
\begin{eqnarray}
x &=& x'/\sqrt{\cos\theta}, \;y = y'\sqrt{\cos\theta}, \;z = z', \\
\partial_x &=& \partial_{x'}\sqrt{\cos\theta}, \;\partial_y = \partial_{y'}/\sqrt{\cos\theta}, \;\partial_z = \partial_{z'},\\
b_x &=& b_{x'}/\sqrt{\cos\theta}, \;b_y = b_{y'}\sqrt{\cos\theta}, \;b_z = b_{z'}.
\end{eqnarray}
Here $\bm b' =\hat \rho b_{\rho '}(\rho',z') + \hat z b_{z'}(\rho',z')= -\nabla' \phi'$ is the transformed magnetic induction within the cylinder, subject to the boundary conditions $b_{z'}(\rho',0) = B_c \cos\theta$ for $\rho' < R/\sqrt{\cos\theta}$, $b_{z'}(\rho',0) = 0$ for $R/\sqrt{\cos\theta} < \rho' < R_0/\sqrt{\cos\theta}$, $b_{z'}(\rho',\infty) = B \cos\theta,$ and  $b_{\rho'}(R_0/\sqrt{\cos\theta},z') = 0$.  This transformed problem can be solved as in Sec.\ \ref{Case1}.

Taking the sum of $F_1$ and $F_2$ and making use of $R/R_0 = \sqrt{f_n}$, we find that the net cost in free energy per unit sample volume for case $1^\prime$ is
\begin{equation}
\Delta F_{1^\prime} = \frac{B_c^2}{\mu_0}\Big[\frac{\delta \sqrt{f_n}}{R_0}+\frac{4  R_0 f_n \cos^{3/2}\!\theta S_1(\sqrt{f_n})}{d}\Big].
\end{equation}

For  fixed values of $f_0$, $\theta_0$, $f_n,$ and $\theta$, we obtain at the value of $R_0$ that minimizes $\Delta F_{1^\prime}$,
\begin{eqnarray}
R_0 &=& \frac{\sqrt{\delta d f_n}}{2\Phi_{1^\prime}}, \label{R0min1''} \\
R &=& \frac{f_n\sqrt{\delta d}}{2\Phi_{1^\prime}}, \label{Rmin1''} \\
\Delta F_{1^\prime} &=& \frac{4B_c^2}{\mu_0}\Big(\frac{\delta}{d}\Big)^{1/2}\Phi_{1^\prime}, \label{DeltaF1''}\\
\Phi_{1^\prime} &=& [\cos^{3/2}\!\theta f_n^{3/2} S_1(\sqrt{f_n})]^{1/2}.\label{Phi1'}
\end{eqnarray}

\subsubsection{Case 2$^\prime$, moderate $f_n$\label{Case2'}}

For intermediate values of   $f_n$ (case 2$^\prime$, extending case 2, but assuming that the domains align along the parallel field component) we approximate the magnetic structure as a periodic array (with period $D$) of  normal domains of width $W= 2R$ parallel to the $y$ axis with periodicity $D = 2R_0$ along the $x$ direction.
The normal domains, containing magnetic flux density $B_c$ (see Fig.\ \ref{BFig}), alternate with flux-free superconducting domains.  The area of normal-superconducting interface per unit sample volume is is $2/D$, and as in Eq.\ (\ref{F12}) the wall-energy cost of the intermediate state per unit sample volume is
\begin{equation}
F_1 = \frac{B_c^2 \delta}{\mu_0 D} = \frac{B_c^2 \delta}{2\mu_0 R_0}. \label{F12'}
\end{equation}

The field-energy cost of the intermediate state per unit sample volume is
\begin{equation}
F_2 = \frac{1}{2 \mu_0 R_0 d} \int dA(b^2-B_0^2\cos^2\theta_0), \label{F22'int}
\end{equation}
where the integral is to be carried out over the area of width $D = 2R_0$ and infinite height above the sample surface at $z = 0$.   Here $\bm b =\hat x b_x(x,z) + \hat z b_z(x,z)= -\nabla \phi$ is the magnetic induction within this area, subject to the boundary conditions $b_z(x,0) = B_c\cos\theta$ for $|x| < R$, $b_z(x,0) = 0$ for $R < |x| < R_0$, and  $b_x(\pm R_0,z) = 0$.  This can be solved as in Sec.\ \ref{Case2}.

Taking the sum of $F_1$ and $F_2$, we find that the net cost in free energy per unit sample volume for case 2$^\prime$ is
\begin{equation}
\Delta F_{2^\prime} = \frac{B_c^2}{\mu_0}\Big[\frac{\delta }{2 R_0}+\frac{2 \cos^2\theta R_0 S_2(f_n)}{d}\Big].
\end{equation}
At the value of $R_0$ that minimizes $\Delta F_{2'}$, we obtain
\begin{eqnarray}
R_0 &=& \frac{\sqrt{\delta d}}{4\Phi_{2'}}, \label{R0min2'} \\
R &=& \frac{f_n\sqrt{\delta d}}{4\Phi_{2'}}, \label{Rmin2'} \\
\Delta F_{2'} &=& \frac{4B_c^2}{\mu_0}\Big(\frac{\delta}{d}\Big)^{1/2}\Phi_{2'}, \label{DeltaF2'}\\
\Phi_{2'} &=& [S_2(f_n)]^{1/2}\cos\theta/2.\label{Phi2'}
\end{eqnarray}

\subsubsection{Case 3$^{\prime}$, large $f_n${\label{Case3'}}}

For large values of $f_n
 = B/B_c$ in an applied field with a parallel component (case 3$^{\prime}$, as an extension of case 3) we approximate the magnetic structure as an array of  identical cylindrical superconducting tubes, oriented parallel to $\hat B$ at angle $\theta$, as shown in Fig.\ \ref{BFig}.  In the plane normal to $\hat B$ the superconducting tubes are assumed to arrange themselves as an equilateral triangular array (with lattice parameter $D$) of circular superconducting tubes of radius  $R$  surrounded by the flux-filled normal phase.  For doing calculations of the magnetic energy of the structure above and below the sample, however, it is convenient to replace the equilateral triangular array inside the sample with a Wigner-Seitz cylinder of radius  $R_0 = (\sqrt{3}/2\pi)^{1/2}D$ with the same  area as the unit cell.
This Wigner-Seitz cylinder is parallel to $\hat B$.
The magnetic flux carried by this Wigner-Seitz cylinder is $\Phi= B\pi R_0^2 = B_c(\pi R_0^2-\pi R^2)$, so that $B/B_c = f_n = 1-f_s$, and $R/R_0 = \sqrt{f_s} = \sqrt{1-f_n}$.  

Within the Wigner-Seitz cylinder of  volume $\pi R_0^2d/\cos\theta$, the area of the normal-superconducting interface is $2\pi R d/\cos\theta$.  From Eq.\ (\ref{gamma}), we see that   the wall-energy cost of the intermediate state per unit sample volume is
\begin{equation}
F_1 =\frac{B_c^2 \delta R}{\mu_0 R_0^2}. \label{F13'}
\end{equation}

The Wigner-Seitz cylinder intersects the top surface at an angle $\theta$, so that the intersection is an ellipse of semimajor axis $R_{0s} = R_0/\cos\theta$ along the $x$ direction and semiminor axis $R_0$ along the $y$ direction.
Similarly, the intersection of the superconducting tube with the surface is  an ellipse of semimajor axis $R_{s} = R/\cos\theta$ along the $x$ direction and semiminor axis $R$ along the $y$ direction.  
The $z$ component of the average magnetic flux density emerging from the Wigner-Seitz ellipse is $B \cos\theta$, but the area of the ellipse is larger by a factor of $1/\cos\theta$ than the cross-sectional area of the Wigner-Seitz cylinder inside the sample, so that the total magnetic flux is the same,  $\Phi= B\pi R_0^2$.

As in Sec.\ \ref{Case1'}, the field-energy cost of the intermediate state per unit sample volume is
\begin{equation}
F_2 = \frac{\cos\theta}{\pi \mu_0 R_0^2 d} \int dV(b^2-B^2\cos^2\theta), \label{F23'int}
\end{equation}
where the integral is to be carried out within a new Wigner-Seitz elliptical cylinder above the sample surface at $z = 0$.  However, this integral can be evaluated using the same scale transformation we used in Sec.\ \ref{Case1'}.

Taking the sum of $F_1$ and the resulting $F_2$, and making use of $R/R_0 = \sqrt{f_s}$, we find that the net cost in free energy per unit sample volume for case ${3^\prime}$ is
\begin{equation}
\Delta F_{3^\prime} = \frac{B_c^2}{\mu_0}\Big[\frac{\delta \sqrt{f_s}}{R_0}+\frac{4  R_0 f_s \cos^{3/2}\!\theta S_1(\sqrt{f_s})}{d}\Big].
\end{equation}

For  fixed values of $f_0$, $\theta_0$, $f_n,$ and $\theta$, we obtain at the value of $R_0$ that minimizes $\Delta F_{3^\prime}$,
\begin{eqnarray}
R_0 &=& \frac{\sqrt{\delta d f_s}}{2\Phi_{3^\prime}}, \label{R0min3'} \\
R &=& \frac{f_s \sqrt{\delta d}}{2\Phi_{3^\prime}}, \label{Rmin3'} \\
\Delta F_{3^\prime} &=& \frac{4B_c^2}{\mu_0}\Big(\frac{\delta}{d}\Big)^{1/2}\Phi_{3^\prime}, \label{DeltaF3'}\\
\Phi_{3^\prime} &=& [\cos^{3/2}\!\theta f_s^{3/2} S_1(\sqrt{f_s})]^{1/2}.\label{Phi3prime}
\end{eqnarray}

\subsubsection{Free-energy comparisons in an inclined field\label{Comps}}

\begin{figure}
\includegraphics[width=8cm]{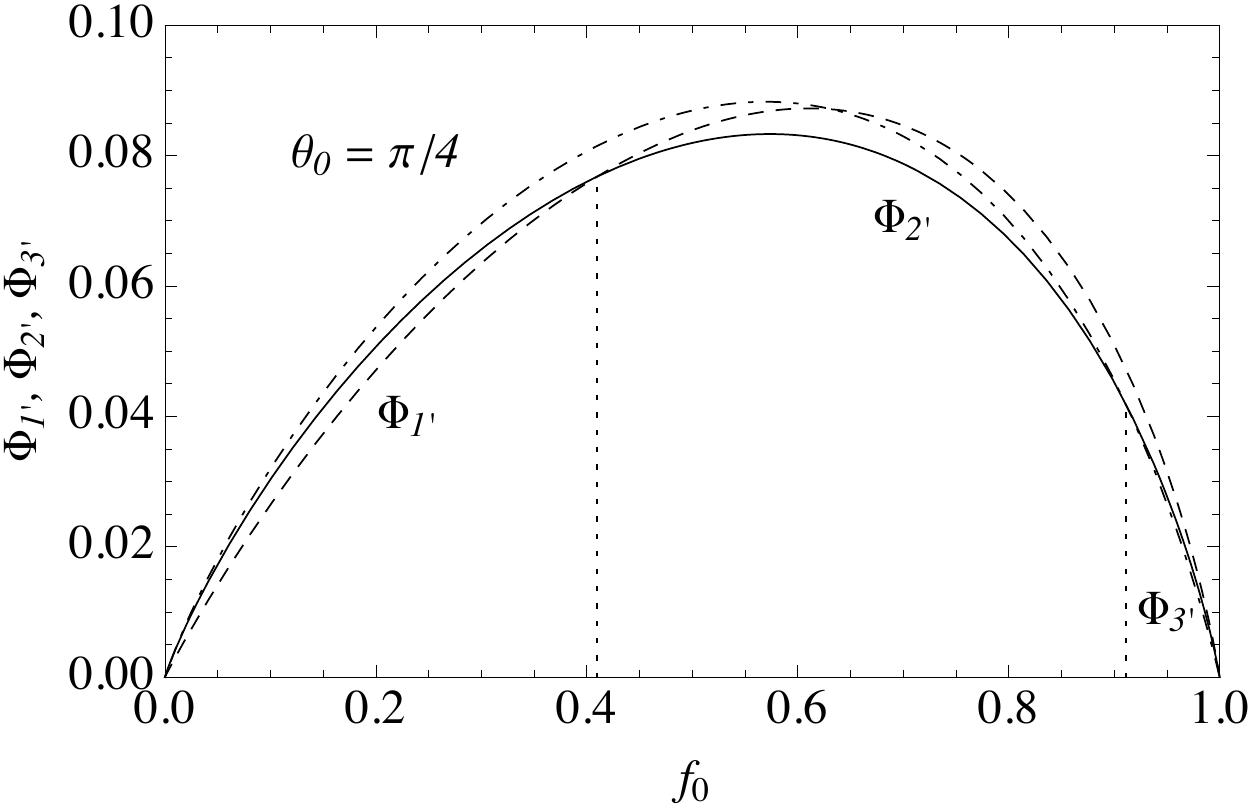}
\caption{%
Plots of the normalized free energies vs $f_0$ in an inclined field with $\theta_0 = \pi/4$ for  case 1$^\prime$: $\Phi_{1^\prime}$ (dashed), array of normal flux tubes, Eq.\ (\ref{Phi1'}); case 2$^\prime$: $\Phi_{2^\prime}$ (solid), parallel array of normal and superconducting domains, Eq.\ (\ref{Phi2'}); and case 3$^\prime$: $\Phi_{3^\prime}$ (dot-dashed), array of superconducting cylinders, Eq.\ (\ref{Phi3prime}).  The free energy cost $\Delta F$ for $\theta_0 = \pi/4$ is smallest for case 1$^\prime$ when $0 \le f_0 \le 0.410$, case 2$^\prime$ for $0.410 \le f_0 \le 0.911$, and case 3$^\prime$ for $0.911 \le f_0 \le 1$.}
\label{PhiPrime123Fig}
\end{figure}

Tilting the applied field away from the normal (i.e., increasing $\theta_0$)   initially increases the range of values of $f_0 = H_0/H_c$ over which the parallel-domain structure is energetically favorable.  An example of this behavior is shown in Fig.\ \ref{PhiPrime123Fig}, in which for $\theta_0 = \pi/4$ the range of values of $f_0$ over which the parallel-domain structure is favored has expanded to $0.410 \le f_0 \le 0.911$ from the range $0.346 \le f_0 \le 0.654$ shown for $\theta_0 = 0$ in Fig.\ \ref{Phi123Fig}.

\begin{figure}
\includegraphics[width=8cm]{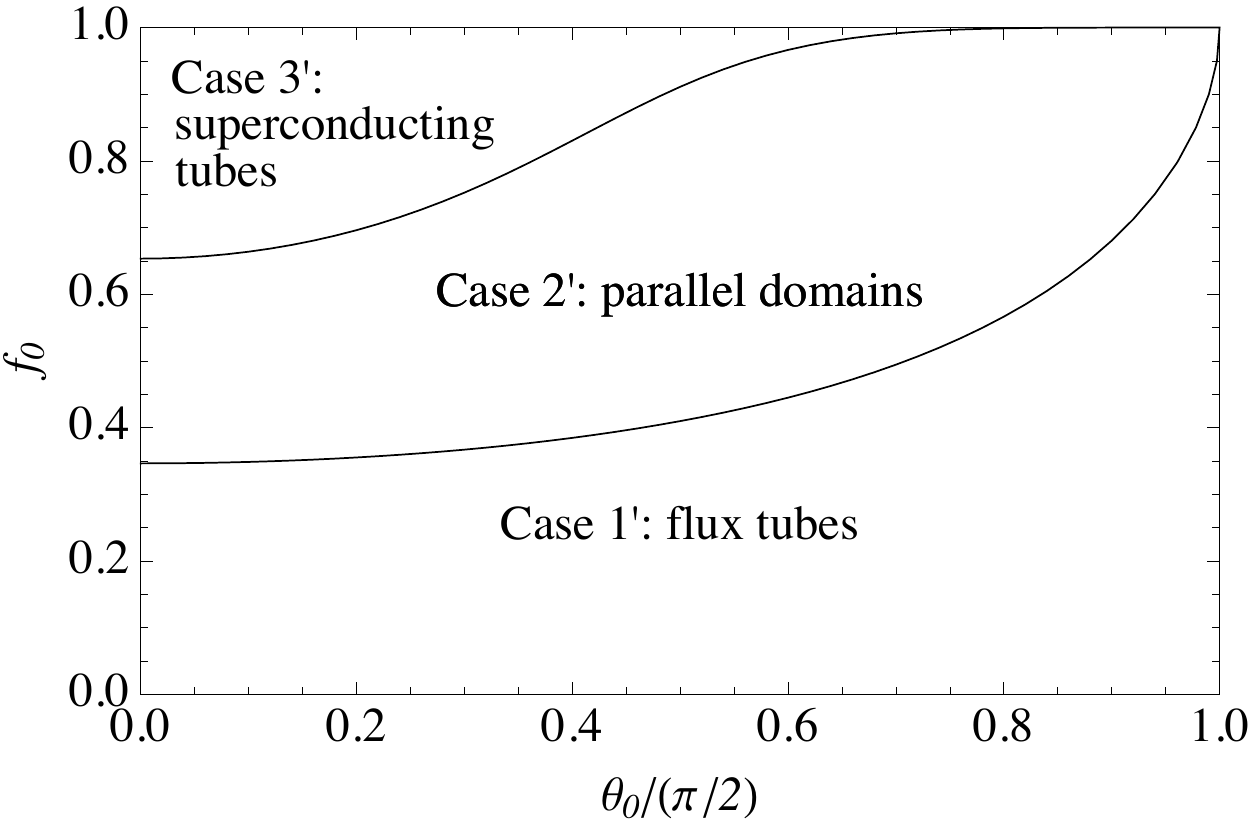}
\caption{%
Diagram indicating which of the intermediate-state structures considered theoretically in Secs.\ \ref{Case1'}-\ref{Case3'} is energetically favored for given values of $f_0 = H_0/H_c$ and $\theta_0$ (see Figs.\ \ref{BFig}, \ref{Phi123Fig}, and \ref{PhiPrime123Fig}). Case 1$^\prime$ ($\Phi_{1^\prime}$) is favored for small $f_0$, case 2$^\prime$ ($\Phi_{2^\prime}$) for intermediate values of $f_0$ and larger values of $\theta_0$, and  case 3$^\prime$ ($\Phi_{3^\prime}$) for large $f_0$ and smaller values of $\theta_0$.
}
\label{f0theta0Fig}
\end{figure}

Figure \ref{f0theta0Fig} shows a diagram indicating which intermediate-state structure is favored for given values of $f_0 = H_0/H_c$ and field angle $\theta_0$ (see Fig.\ \ref{BFig}).  Parallel domains (case 2$^\prime$, the Sharvin structure\cite{Sharvin57}) are favored for $f_0$ not far from 1 when the tilted field $\bm H_0$ is nearly parallel to the sample's surface (i.e., when $\theta_0$ is not far from $\pi/2$).

\section{Summary and conclusions\label{Concl}}
In  summary, we have compared the energies of three realistic topologies of the intermediate state in an infinite type-I superconducting slab of macroscopic thickness. The present calculation takes the energy of the external field for all three cases into account in the same  manner, which is a significant improvement, since the energies of the different topologies are so close to each other. Specifically, we use essentially the same method to calculate the free energy of the intermediate-state structure as a function of the applied field for three assumed model structures, described briefly as model (1) an array of normal flux tubes surrounded by superconducting phase, model (2) an array of alternating parallel normal and superconducting domains, and model (3) an array of superconducting tubes surrounded by flux-filled normal phase.  We find that as the applied field increases, the structures with the lowest free energies are, in order,  (1), (2), and (3).  However, magneto-optical images of the intermediate state in the corresponding field ranges are generally not accurately described by one of these three simple  models, and there are many reasons for this, 
as detailed below.

Nevertheless, the images tend to show that the intermediate-state structure most closely resembles model (1) (separated normal domains) at low applied fields, model (2) (stripe-like, connected normal domains) at intermediate fields, and model (3) (separated superconducting domains) at high fields.

It is found that for a close to perpendicularly applied field, superconducting tubes (at high field), stripes (at intermediate field) and normal flux tubes or `macro vortices' occupy roughly equal ranges in field from zero to the critical value, $B_c$. For a close to in-plane applied field, for all but the highest fields, normal tubes have the lowest energy. 
We note that practically all published magneto-optical images for nearly in-plane applied field (Sharvin geometry) show a very nice \textit{laminar} pattern. The reason for this apparent discrepancy is that these experiments were generally done close to $H_c$, i.e. for $f_0\simeq 1$. These experiments are thus consistent with our theoretical result. A systematic experimental exploration of the phase diagram in the region 
$\theta_0/\left( \pi /2\right) \gtrsim 0.9$ for the whole range 
$0\leq f_0\leq 1$ would be certainly interesting.

An important extension of the present work would take into account the deviation from $B_c$ of the magnetic flux density in the normal domains. This is in particular important for samples with thickness comparable to or smaller than $\lambda$, $\xi$, or $\delta$. For this, an extra contribution to the free energy density must be taken into account. In a recent work \cite{Kozhevnikov2012} this was done for the laminar pattern only.
Furthermore, corrugations could in principle be taken into account, as was shown by Faber \cite{Faber58}. See also the upper image in our Fig.\ \ref{diagram}. However, the analytical calculation of the field energy would become much more complicated, if possible at all.

As noticed in the introduction, we reiterate that experimental patterns may be quite different from the thermodynamic predictions discussed in this paper because of several factors. The small energy difference between different patterns in conjunction with the effect of residual pinning remaining in carefully annealed samples may add to hysteretic effects. Additionally, since the tubes repel each other at large distances, there are local barriers for the transformation from tubular (closed topology) to laminar (open topology) patterns \cite{Huebener1974,Tinkham1996}. Also, many experiments were conducted on well - controlled and well - characterized thin films, which, however, do not obey our starting assumption of the thick slab with its thickness greater than all of the characteristic length scales. In addition, flux tubes become more favorable for larger values of Ginzburg-Landau parameter \cite{Lukyanchuk2001} and for very thin type-I superconductors where the pattern turns into Abrikosov vortices \cite{Tinkham1996}. Furthermore, there is a possibility of quantum tunneling of the domain S/N walls \cite{Chudnovsky2011} as well as a pronounced effect of confined geometry. Recent numerical and experimental results obtained on mesoscopic samples show the tendency toward increased stability of flux tubes \cite{Simonin1980,Hernandez2005,Berdiyorov2009,Berdiyorov2012}.

We conclude by stating that the structure of the intermediate-state of type-I superconductors is remarkably complicated, and no single theoretical paper would be able to provide an accurate description of all cases. Yet, we believe it is important to have a general thermodynamic picture, which is what we offer here.

\begin{acknowledgments}
This research was supported  by the Department of Energy - Basic Energy Sciences under Contract No. DE-AC027CH11358  and by the Center for Emergent Superconductivity, an Energy Frontier Research Center funded by the U.S. Department of Energy, Office of Science, Office of Basic Energy Sciences under Award Number DE-AC0298CH1088.  We thank V. G. Kogan for helpful discussions.

\end{acknowledgments}


\end{document}